\documentclass[aps,prl,twocolumn,preprintnumbers]{revtex4}

\usepackage{graphicx}
\usepackage{dcolumn}
\usepackage{bm}
\usepackage{amsmath}
\usepackage{amssymb}
\usepackage{amsfonts}
\usepackage{float}
\usepackage{hyperref}[11in]
\usepackage{dsfont}
\usepackage{slashed}
\usepackage{booktabs}
\usepackage{multirow}
\usepackage{subfigure}
\usepackage[sort&compress]{natbib}
\usepackage{xcolor}
\usepackage{color}
\usepackage{colordvi}
\usepackage{wasysym}

\usepackage{graphicx}
\usepackage{xfrac}

\newcommand{\be}{\begin{equation}}
\newcommand{\ee}{\end{equation}}
\newcommand{\beq}{\begin{eqnarray}}
\newcommand{\eeq}{\end{eqnarray}}

\newcommand{\tr}{\mathrm{Tr}}

\newcommand{\<}{\langle}
\renewcommand{\>}{\rangle}

\begin{document}
\title{Nucleon strange electromagnetic form factors}

\author{
  C.~Alexandrou$^{1,2}$,
  S.~Bacchio$^{2}$,
  M.~Constantinou$^{3}$,
  J. Finkenrath$^{2}$,
  K.~Hadjiyiannakou$^{2}$,
  K.~Jansen$^{4}$,
  G.~Koutsou$^{2}$
    \\(Extended Twisted Mass Collaboration)
}
\affiliation{
  $^1$Department of Physics, University of Cyprus, P.O. Box 20537, 1678 Nicosia, Cyprus\\
  $^2$Computation-based Science and Technology Research Center, The Cyprus Institute, 20 Kavafi Str., Nicosia 2121, Cyprus \\
  $^3$Department of Physics, Temple University, 1925 N. 12th Street, Philadelphia, PA 19122-1801,
USA\\
  $^4$NIC, DESY, Platanenallee 6, D-15738 Zeuthen, Germany\\
}

\begin{abstract}
  \centerline{\includegraphics[width=0.15\linewidth]{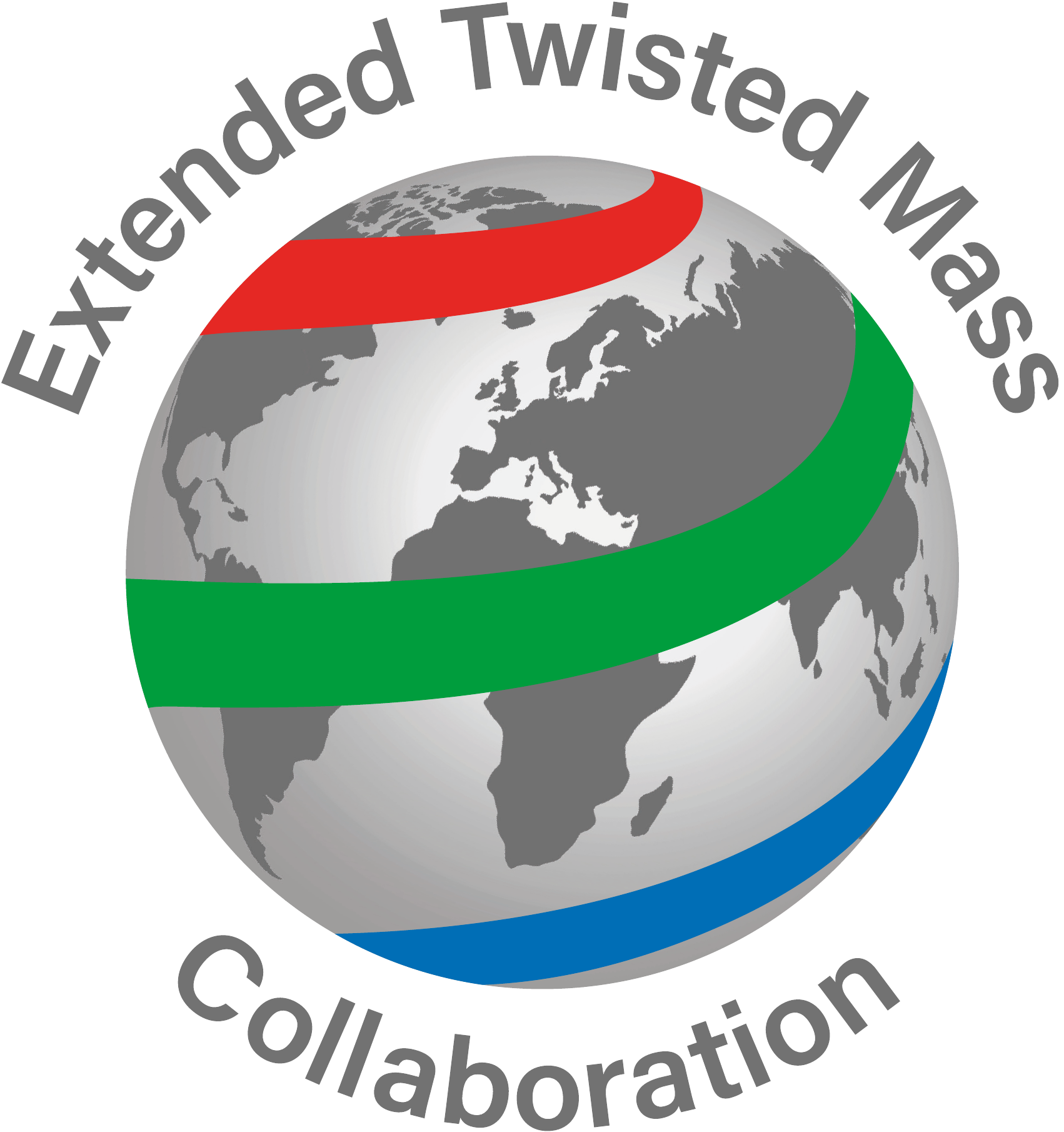}}
  \vspace*{0.3cm}
  The role of the strange quarks on the low-energy interactions of the proton  can be probed through the strange electromagnetic form factors.  Knowledge of these form factors provides essential input for  parity-violating processes and contributes to the understanding of the sea quark dynamics.  We determine
  the strange electromagnetic form factors of the nucleon within the lattice formulation of Quantum Chromodynamics  using simulations that include light, strange and charm quarks in the sea all tuned to their physical mass values.
  We employ state-of-the-art techniques to accurately  extract the form factors for values of the momentum transfer square 
up to 0.8~GeV$^2$. We find that both the electric and magnetic form factors are statistically non-zero. We obtain for the strange magnetic moment $\mu^s=-0.017(4)$, 
the strange magnetic radius $\langle r^2_M \rangle^s=-0.015(9)$~fm$^2$, and the strange charge
radius $\langle r^2_E \rangle^s=-0.0048(6)$~fm$^2$.
\end{abstract}

\maketitle
\bibliographystyle{apsrev}

{\it{Introduction:}} Strange quarks are the lightest non-valance quarks in the nucleon and thus the most likely constituents to contribute to sea-quark dynamics.  The  study of strange-quark contributions  to nucleon
structure observables allows to uniquely identify sea-quark effects and understand virtual particle dynamics in the non-perturbative regime of Quantum Chromodynamics (QCD). 
A possible difference in the spatial distribution of strange and anti-strange quarks 
reflected by a non-zero strange electric form factor $G_E^s(Q^2)$, and a finite strange
magnetic moment $\mu^s\equiv G_M^s(Q^2{=}0)$ are key quantities describing the non-trivial composite structure of the nucleon.  Parity violating electron-proton elastic scattering  
events probing the interference of photons and Z-bosons exchanges enable the measurement of the strange 
form factors and weak charge of the proton. An accurate determination of the neutral-weak vector
form factor  in combination with the electromagnetic form factors of the nucleon  are needed in order to put constraints on new physics beyond the standard model (SM). 

 A number of major experiments have been measuring the parity violating form factors of the proton  seeking to detect beyond the SM physics. The experimental program to  study the strangeness in the proton has a long history beginning with the SAMPLE experiment~\cite{Spayde:2003nr,Beise:2004py} and continuing with the series of A4 experiments at the Mainz Microtron accelerator facility (MAMI)~\cite{Maas:2004ta,Maas:2004dh,Baunack:2009gy}  and the HAPPEX~\cite{Aniol:2005zf,Aniol:2005zg,Acha:2006my,Ahmed:2011vp} and G0 experiments~\cite{Armstrong:2005hs,Androic:2009aa} at JLab. However to date, the experimental results, although indicating non-zero values, carry large errors that make them inconclusive. This is confirmed by a recent global analysis of parity-violating elastic scattering data~\cite{Gonzalez-Jimenez:2014bia}, where although a negative magnetic strange form factor is indicated, the large error still makes it consistent with zero.
 A review of the  experimental program and results can be found in 
Ref.~\cite{Maas:2017snj}. In addition, a number of phenomenological studies have been devoted to the study of the strangeness in the proton~\cite{Weigel:1995jc,Lyubovitskij:2002ng,Silva:2005qm,Goeke:2006gi,Bijker:2005pe,Wang:2013cfp,Hobbs:2014lea}.

Given the current status of the  experimental results, where there is no agreement  even  on the sign of the strange electromagnetic 
form factors, a first principle calculation of these key quantities is crucial.
 Lattice QCD provides
 a rigorous framework to compute non-perturbatively these quantities. However, it is only recently that efficient algorithms enable us to simulate the theory
 with physical values of the light quark masses and to evaluate disconnected 
 quark loops to sufficient accuracy~\cite{Green:2015wqa,Sufian:2016pex,Sufian:2017osl,Alexandrou:2018zdf}. In this work, we use simulations generated with physical values of the light quark masses to
 evaluate accurately the strange quark loops and extract  the electromagnetic
 strange form factors directly at the physical point.

{\it{Lattice methodology:}} The final results of this work are based on the analysis of an ensemble simulated with two mass degenerate light
quarks, a strange and a charm quark ($N_f=2+1+1$)  with  masses tuned to their
physical values~\cite{Alexandrou:2018egz}. We use the twisted mass formulation~\cite{Frezzotti:2000nk,Frezzotti:2003ni,Frezzotti:2004wz} including a clover term~\cite{Sheikholeslami:1985ij} for the simulations.
The lattice volume is $ 64^3 \times 128$, $m_\pi L = 3.62$, where $L$ is the spatial lattice length and the pion mass $m_\pi=0.1393(7)$~MeV
and the lattice spacing $a=0.0801(4)$~fm determined 
from the nucleon mass~\cite{Alexandrou:2018sjm}. We will refer to this ensemble as the cB211.072.64 ensemble.
We use Osterwalder-Seiler strange and charm quarks with mass  tuned to reproduce the $\Omega^-$ baryon mass and the mass of $\Lambda_c$ respectively.

The nucleon matrix element of the electromagnetic operator
 decomposes into two CP-even form factors given by
\be
\< N(p',s') \vert j_\mu^f \vert N(p,s) \> \propto \bar{u}_N(p',s') \Lambda^f_\mu(q^2) u_N(p,s)
\label{Eq:VMEdecomp}
\ee
with
\be
\Lambda^f_\mu(q^2) = \gamma_\mu F_1^f(q^2) + \frac{i \sigma_{\mu\nu} q^\nu}{2 m_N} F_2^f(q^2),
\label{Eq:F1F2}
\ee
where  $F_1^f(q^2)$,  $F_2^f(q^2)$ are the Dirac and Pauli form factors with the upper index ``$f$" indicating quark flavors.
$N(p, s)$ is the nucleon state with initial (final) momentum
$p(p')$ and spin $s(s')$, with energy $E_N(\vec{p})$ ($E_N(\vec{p}\,')$)
and mass $m_N$. The momentum transfer squared is $q^2=q_\mu q^\mu$ where
$q_\mu = (p_\mu' - p_\mu)$ and $u_N$ is the nucleon spinor. 
Since we are interested in the strange quark contributions we take 
$j_\mu^s = e_s\bar{s}(x) \gamma_\mu s(x)$, 
where $e_s=-1/3$.
The electric and magnetic Sachs form factors can be expressed
as linear combinations of the Dirac and Pauli form factors given in Euclidean space via
the relations,
\beq
G_E^s(Q^2) &=& F_1^s(Q^2) - \frac{Q^2}{4 m_N^2} F_2^s(Q^2) \\
G_M^s(Q^2) &=& F_1^s(Q^2) + F_2^s(Q^2)
\label{Eq:GEGM}
\eeq
where $Q^2=-q^2$.
In order to extract the electric and magnetic strange form factors $G_E^s$ and $G_M^s$
in lattice QCD we need the evaluation of two- and three-point
correlation functions, given by
\be
C(\Gamma_0,\vec{p};t_s) = \sum_{\vec{x}_s} \tr \left[ \Gamma_0 \< J_N(t_s,\vec{x}_s) \bar{J}_N(0,\vec{0})\>\right] e^{-i\vec{x}_s \cdot \vec{p}} 
\label{Eq:2pt}
\ee
\beq
&&C_\mu^s(\Gamma_\nu,\vec{q},\vec{p}\;';t_s,t_{\rm ins}) = \sum_{\vec{x}_s,\vec{x}_{\rm ins}} e^{+i\vec{x}_{\rm ins}\cdot\vec{q} - i \vec{x}_s \cdot \vec{p}\,'} \times \nonumber\\ 
&& \tr \left[\Gamma_\nu \< J_N(t_s,\vec{x}_s) j^s_\mu(t_{\rm ins},\vec{x}_{\rm ins}) \bar{J}_N(0,\vec{0})\> \right],
\label{Eq:3pt}
\eeq
where  $x_0=(0,\vec{0})$ is the position at which the nucleon is created (source),  $x_s$   the lattice point at which the nucleon is annihilated (sink) and   $x_{\rm ins}$  denotes the lattice site at which the current couples to a quark.
We use projectors $\Gamma_\nu$ taking for the unpolarized $\Gamma_0 = \frac{1}{2} (1+\gamma_0)$ and for the polarized  $\Gamma_k=\Gamma_0 i \gamma_5 \gamma_k$.
$J_N(x)=\epsilon^{abc} u^a(x) [u^{bT}(x)\; \mathcal{C} \gamma_5\; d^c(x)]$ is the standard interpolating field of the nucleon where $u,d$ are the up/down quark fields and $\mathcal{C}$ is the charge conjugation matrix. We use Gaussian smearing \cite{Alexandrou:1992ti,Gusken:1989qx} for the quark interpolating fields
with APE-smeared \cite{Albanese:1987ds} gauge links in the hopping operator in order to
increase the overlap with the ground state. The parameters are optimized to yield a nucleon mean square radius of about 0.5 fm which ensures an early plateau in the two-point nucleon correlator~\cite{Alexandrou:2018sjm}.

The electromagnetic strange current $j^s$ necessarily couples to a vacuum strange quark. The contribution of the strange quark loop is given by 
\be
\sum_{\vec{x}_{\rm ins}} e^{+i\vec{q}\cdot \vec{x}_{\rm ins}}\tr[ \gamma^\mu \;G(x_{\rm ins};x_{\rm ins})].
\label{Eq:loop}
\ee
To evaluate Eq.(\ref{Eq:loop}) we need to  compute the sum over the  diagonal  spatial 
components of the strange quark  propagator $G(x;y)$ that would require $12 \times L^3$ inversions currently not feasible.
Therefore, stochastic approaches~\cite{Wilcox:1999ab} combined  with dilution methods~\cite{Foley:2005ac} are employed to compute such quark loops~\cite{Alexandrou:2019ali}. 
We perform a full dilution in spin and color~\cite{Alexandrou:2012zz} to avoid any
stochastic contamination in that subspace. The elements of the propagator
decay exponentially with the distance $\vert x - y \vert$ thus dilution in
space-time up to a specific distance 
reduces stochastic contamination entering from off-diagonal elements. This is implemented by employing the hierarchical probing technique~\cite{Stathopoulos:2013aci} using a four dimensional coloring of distance-$2^3$
resulting in $N_{Had}=512$ Hadamard vectors. The coloring ensures an exact cancellation of up
to 4 closest neighbors to the diagonal.
We also exploit properties of
the twisted mass fermions in the so-called one-end trick~\cite{Michael:2007vn,McNeile:2006bz}, which provides an increased noise-to-signal ratio~\cite{Alexandrou:2013wca,Abdel-Rehim:2013wlz}. We compute the quark loops for
every time-slice $t_{\rm ins}$ and  construct  the disconnected three-point function at every value of  $t_s$ and $t_{\rm ins}$ by correlating it with  
 200 nucleon two-point functions  using randomly 
distributed source positions per gauge configuration. We use  750
configurations, averaging over proton and neutron and forward and backward propagators to reach in total
600,000 measurements. 
We utilize the multi-grid algorithm implemented on GPUs through
the QUDA software \cite{Babich:2010mu,Clark:2016rdz,Alexandrou:2014fva} to accelerate the calculation of the quark propagators.

The  nucleon matrix element is extracted from an optimally constructed ratio~\cite{Alexandrou:2013joa,Alexandrou:2011db,Alexandrou:2006ru} given by,
\begin{eqnarray}
&&  R_\mu(\Gamma_\nu,\vec{p}\,',\vec{p};t_s,t_{\rm ins}) = \frac{C_\mu(\Gamma_\nu,\vec{p}\,',\vec{p};t_s,t_{\rm ins})}{C(\Gamma_0,\vec{p}\,'\
;t_s)} \times \nonumber \\
&&  \sqrt{\frac{C(\Gamma_0,\vec{p};t_s-t_{\rm ins}) C(\Gamma_0,\vec{p}\,';t_{\rm ins}) C(\Gamma_0,\vec{p}\,';t_s)}{C(\Gamma_0,\vec{p}\,';t_s\
-t_{\rm ins}) C(\Gamma_0,\vec{p};t_{\rm ins}) C(\Gamma_0,\vec{p};t_s)}} \,,
\label{Eq:ratio}
\end{eqnarray}
which becomes time-independent for $\Delta E (t_s-t_{\rm ins}) \gg 1$ and $\Delta E t_{\rm ins} \gg 1$ yielding  $\Pi_\mu(\Gamma_\nu,\vec{p}\,',\vec{p})$, where $\Delta E$ is the energy gap between the ground and the first excited state. In practice, one  needs to identify the shortest time separation for which the excited states are sufficiently suppressed. We employ three methods to check the
convergence to the ground state matrix element~\cite{Alexandrou:2019ali,Alexandrou:2019brg}: i) \emph{Plateau method:} We use the ratio of Eq.(\ref{Eq:ratio}) and
identify a time-independent window (plateau) where we fit to a constant. We seek convergence of this value as we increase $t_s$; ii) \emph{Two-state fit:} Takes into account the first excited state in the three- and two-point correlators entering in the ratio of Eq.~(\ref{Eq:ratio}) ; iii) \emph{Summation method:}~\cite{Martinelli:1987zd} Summing over the insertion time $t_{\rm ins}$ in Eq.(\ref{Eq:ratio}), excluding contact terms,  leads
to a linear behavior from where the slope yields the nucleon state matrix element as $t_s$ increases.
 For all the three methods we use the correlated $\chi^2$ fits that take into account correlations between different insertion time slices and source-sink time separations.

There are several combinations of $\Pi_\mu(\Gamma_\nu,\vec{p}\,',\vec{p})$ from where
$G_E(Q^2)$ and $G_M(Q^2)$ can be extracted as described in  Ref.~\cite{Alexandrou:2018sjm}. These lead to an over-constrained system of equations
$\Pi = D \; G$, where the form factors $G$ are extracted through a singular value 
decomposition of the coefficients $D$. For disconnected quantities we are not limited to use $\vec{p}\,'=\vec{0}$ since no additional inversions are
needed. Given that $\vec{p}=\frac{2\pi}{L}\vec{n}$, we analyze three-point
functions with $|\vec{n}\,'|^2 \leq 2$, and $|\vec{n}|^2 \leq 11$ for
$G_E$ and $|\vec{n}|^2 \leq 26$ for $G_M$ since the latter is in general more accurate allowing to reach higher values of the momentum.
\begin{figure}[ht!]
  \includegraphics[scale=0.49]{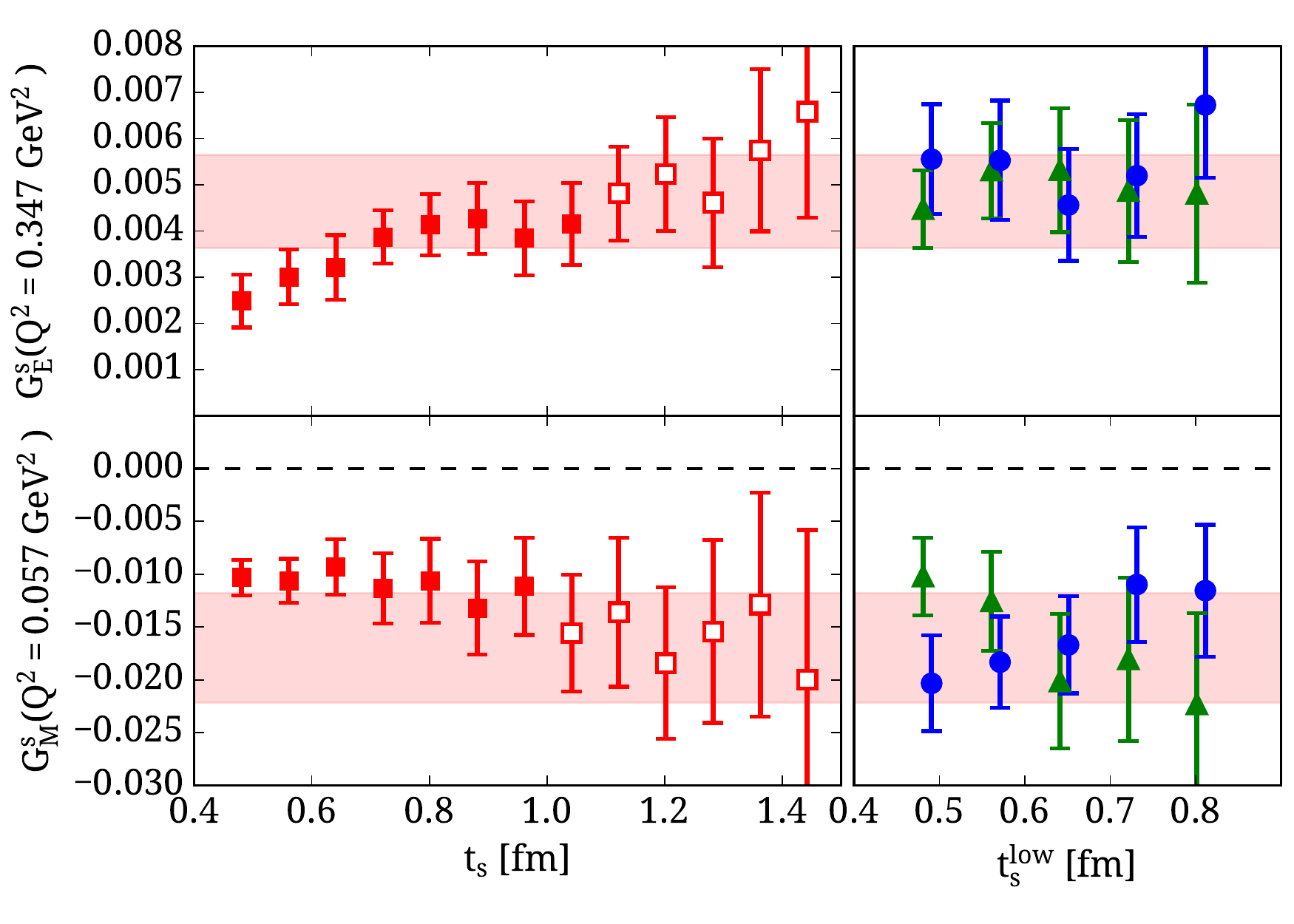}  \\
  \caption{The  electric  $G_E^s(Q^2)$ (upper) and magnetic  $G_M^s(Q^2)$ (lower) form factors  for $Q^2=0.347$~GeV$^2$
  and $Q^2=0.057$~GeV$^2$ respectively. In the
  left panel we show the extracted values using the plateau method as
  a function of $t_s$ (red squares). In the right
  panel we show the extracted values using the summation method (green triangles) and two-state fit (blue circles) as a function of the lowest value of the sink-source time separation, $t_s^{\rm low}$. The largest value of $t_s$  is fixed at $t_s=1.44$~fm for the summation method and at
   $t_s=1.12$~fm for the two-state fit. Open symbols indicate the values of $t_s$ where convergence to the ground is reached. The weighted average of plateau values to extract our final value of the form factor is shown by the red band.}
  \label{Fig:GEPLTSUMM}
\end{figure}

In Fig.~\ref{Fig:GEPLTSUMM} we show the electric and magnetic form 
factors for two representative values of the  momentum transfer squared. As can be seen,  the
 plateau method yields results that are in agreement as $t_s$ is increased with the
two-state fit and summation method. We then perform a weighted average over the plateau values in order to extract the final value. A similar behavior is observed for the other $Q^2$ values. 

{\it{Renormalization:}}
Since we use the local electromagnetic current we need to compute the renormalization function $Z_V$, which is scheme and scale independent simplifying the renormalization procedure. 
  We employ the Rome-Southampton method (RI$'$-MOM scheme)~\cite{Martinelli:1994ty} and use the momentum source approach introduced in Ref.~\cite{Gockeler:1998ye} to achieve per mil statistical accuracy using ${\cal O}(10)$ configurations~\cite{Alexandrou:2010me,Alexandrou:2012mt,Alexandrou:2015sea}. Discretization effects are suppressed using momenta that have the same spatial components, satisfying $ {\sum_i p_i^4}/{(\sum_i p_i^2 )^2}{<}0.3$~\cite{Constantinou:2010gr}. Furthermore, we  subtract unwanted finite-$a$ effects to ${\cal O}(g^2 a^\infty)$ using results from lattice perturbation theory~\cite{Alexandrou:2015sea}.  This procedure is performed using five $N_f{=}4$ ensembles simulated with a range of  pion masses in order to take the chiral limit. These gauge configurations are dedicatedly produced for the renormalization program using  the same $\beta$ value as the $N_f=2+1+1$ ensemble of this work.
 On each ensemble we compute  25 different values of the initial renormalization scale $(a\,\mu_0)^2 \, \in\,  [1{-}7]$. The dependence of $Z_V$ on the pion mass is very mild as confirmed by the fact that the coefficient of the quadratic term in $m_\pi$  is compatible with zero.  After extrapolating to the chiral limit for each 25 values we then extrapolate to  $(a \mu_0)^2 \rightarrow 0$ to remove any residual dependence on the RI$'$-MOM scale. Due to the subtraction of the ${\cal O}(g^2 a^\infty)$ artifacts, an almost constant line is obtained for $Z_V$ for $(a\,\mu_0)^2 \, \in\,  [2{-}7]$. 
We obtain as our final value of  $Z_V=0.728(1)(4)$, where the first parenthesis gives the  statistical error and
 the second the systematic coming from varying the fit window in the $(a \mu_0)^2 \rightarrow 0$ extrapolation. More details of the procedure can be found in Refs.~\cite{Alexandrou:2019ali,Constantinou:2014fka,Alexandrou:2015sea}. 

\begin{figure}[ht!]
  \includegraphics[scale=0.5]{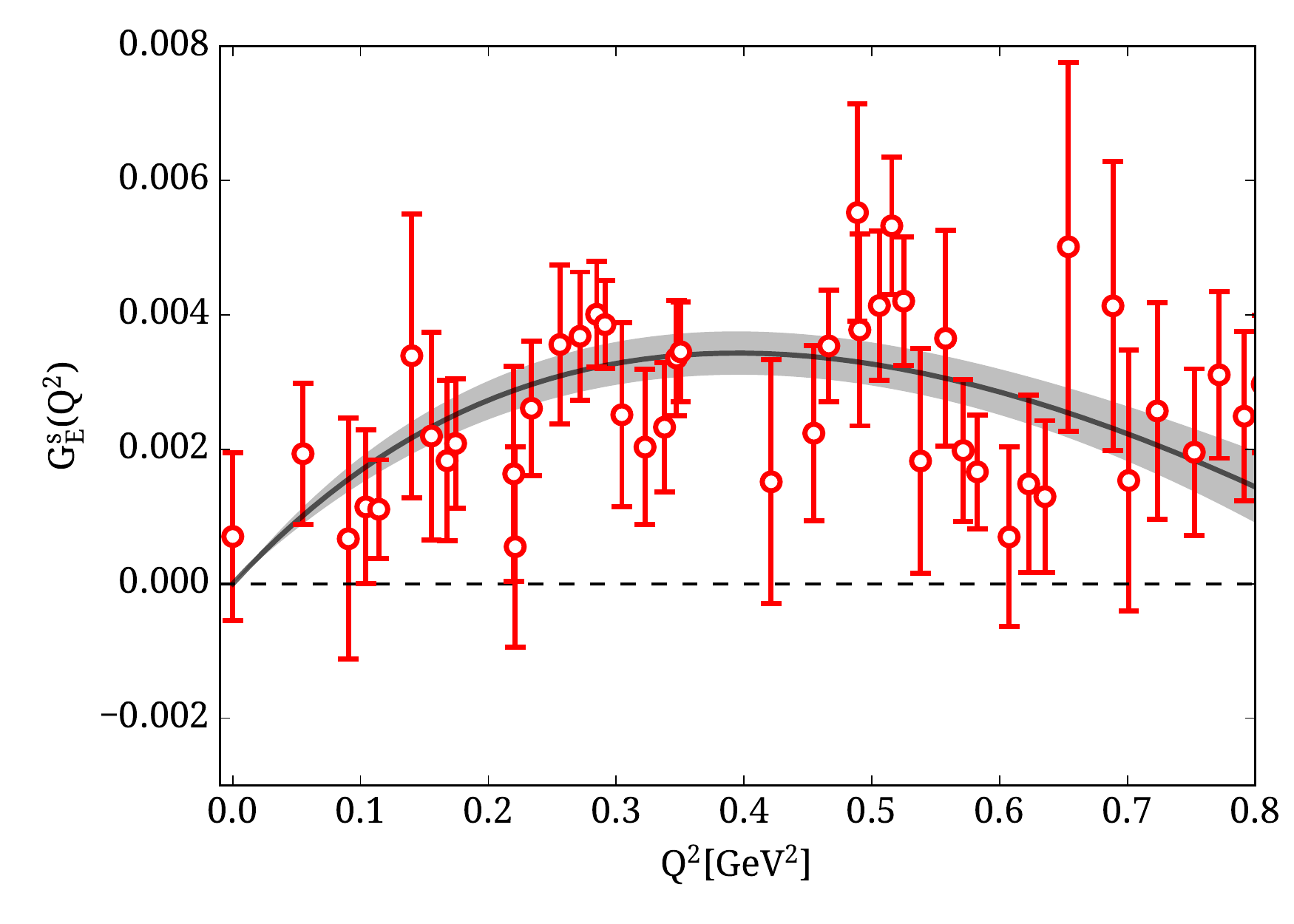}  \\
  \caption{The strange  nucleon electric form factor $G_E^s(Q^2)$ as a function of $Q^2$. The band shows a fit to the form factor using the z-expansion that yields  $\chi^2$/d.o.f=0.94. The strange charge factor $e_s=-1/3$ is not included.}   \label{Fig:GE}
\end{figure}

\begin{figure}[ht!]
  \includegraphics[scale=0.5]{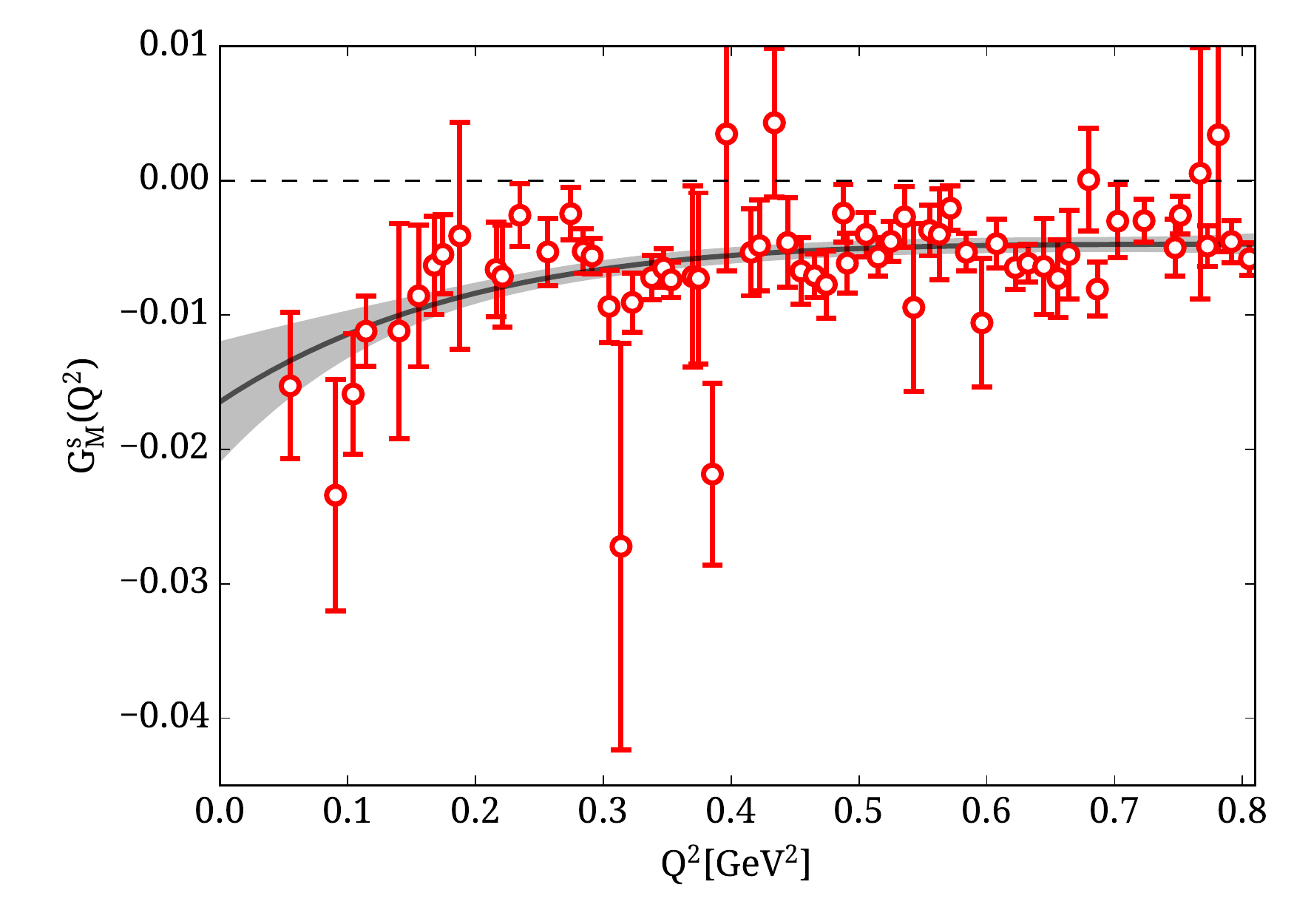} 
  \caption{The nucleon strange magnetic form factor $G_M^s(Q^2)$ as a function of  $Q^2$. The notation is as in Fig.~\ref{Fig:GE}. The fit yields  $\chi^2$/d.o.f=1.05.}
  \label{Fig:GM}
\end{figure}

{\it{Results:}} The results for the strange electric form factor $G_E^s(Q^2)$ are presented in Fig.~\ref{Fig:GE}.  The form factor is zero at $Q^2=0$ as expected and reaches a maximum at about $Q^2\simeq0.4$~GeV$^2$.
In Fig.~\ref{Fig:GM} 
we show results for the strange magnetic form factor $G_M^s(Q^2)$, which is clearly negative and non-zero becoming increasingly more negative as $Q^2\rightarrow 0$.
We fit the $Q^2$ dependence of the form factors employing  the model independent z-expansion~\cite{Bhattacharya:2011ah,Hill:2011wy,Hill:2010yb},
\be
G(Q^2) = \sum_{k=0}^{k_{\rm max}} a_k z^k,\;\;\;\; z(Q^2)=\frac{\sqrt{t_{\rm cut} +Q^2} - \sqrt{t_{\rm cut}}}{\sqrt{t_{\rm cut} +Q^2} - \sqrt{t_{\rm cut}}}\,,
\ee
using as $t_{cut}=(2 m_K)^2$ where $m_K=486(4)$~MeV the kaon mass as measured in this ensemble. 
Since the series is expected to
converge one can truncate to a $k=k_{\rm max}$ and check convergence by increasing $k_{\rm max}$. For the electric form factor
since $G_E^s(0)=0$ we set $a_0=0$. We truncate the series to $k_{\rm max}=5$ since including  higher order terms has an insignificant effect on the fit. In order to stabilize the fit we use Gaussian priors  for the coefficients with $k>1$. Namely we set $a_{k>1}  = 0 \pm w \max(|a_0|,|a_1|)$,
where $w$ is the width of the Gaussian prior. We find that for $w\ge 10$  the extracted  values are unaffected and  therefore we set $w=10$ in the fit. We use a correlated $\chi^2$ fit since the various  $Q^2$ values are correlated. This improves the quality of  the fit  and the extracted values. From the fits we determine the strange
magnetic moment given by the fit parameter $a_0^M$, which is the value of $\mu^s \equiv G_M^s(0)$. The radii are extracted from the slope of the form factors as $Q^2\rightarrow 0$, namely via

\be
\langle r^2_{E,M}\rangle^s  =  -6 \frac{dG^s_{E,M}(Q^2)}{dQ^2} \Bigg \vert_{Q^2=0}
= \frac{-3a_1^{E,M}}{2 t_{\rm cut}}.
\ee
The extracted values are
$\<r^2_E\>^s = -0.0048(6)\,{\rm fm}^2$,
$\<r^2_M\>^s = -0.015(9)\,{\rm fm}^2$ and
$\mu^s = -0.017(4)$, where the error is purely statistical.

We perform the same analysis for the charm electromagnetic form factors. The electric charm form factor $G_E^c$ is consistent with zero while the magnetic $G_M^c$ tends to be negative albeit with large statistical errors that do not exclude zero for most $Q^2$ values. At the lowest available $Q^2$ value we find $G_M^c(Q^2\simeq0.051$GeV$^2)=-0.004(2)$. 

{\it{Comparison:}} Within the twisted mass formulation we have previously analyzed an $N_f=2$ ensemble with close to physical pion mass, namely $m_\pi=130$~MeV, lattice spacing $a=0.094(1)$~fm and lattice size $48^3\times 96$~\cite{Alexandrou:2018zdf}, referred to as the cA2.09.48 ensemble. However, our current analysis yields results with higher accuracy. Besides these two analyses, currently there are no other lattice QCD calculations of these form factors  directly at the physical pion mass. The fact that we achieved the current accuracy is due to our improved methods for computing the quark loops leading to about four times more accurate results. Three other groups have computed the strange form factors with the $\chi$QCD collaboration including an ensemble with close to physical pion mass.
The analysis was performed  using a mixed setup with
$N_f=2+1$ gauge configurations produced using domain wall fermions and 
overlap fermions used for the evaluation of nucleon two- and  three-point correlators. The  four ensembles spanned  pion masses $m_\pi \in [139-330]$~MeV. Their final
values are extracted using a chiral extrapolation since their results at the physical point alone are not accurate~\cite{Sufian:2016pex}.
The other two groups used simulations with heavier than physical pions: The LHPC collaboration analyzed one ensemble of $N_f=2+1$ clover-improved Wilson fermions with $m_\pi=317$~MeV~\cite{Green:2015wqa} and used an interpolation to estimate the value  at the physical point; The third group~\cite{Djukanovic:2019jtp} analyzed several CLS
ensembles of $N_f=2+1$ $\mathcal{O}(a)$-improved Wilson fermions with 
pion masses $m_\pi \in [200-360]$~MeV and performed a chiral extrapolation to extract the value at the physical point.

\begin{figure}[ht!]
  \includegraphics[scale=0.55]{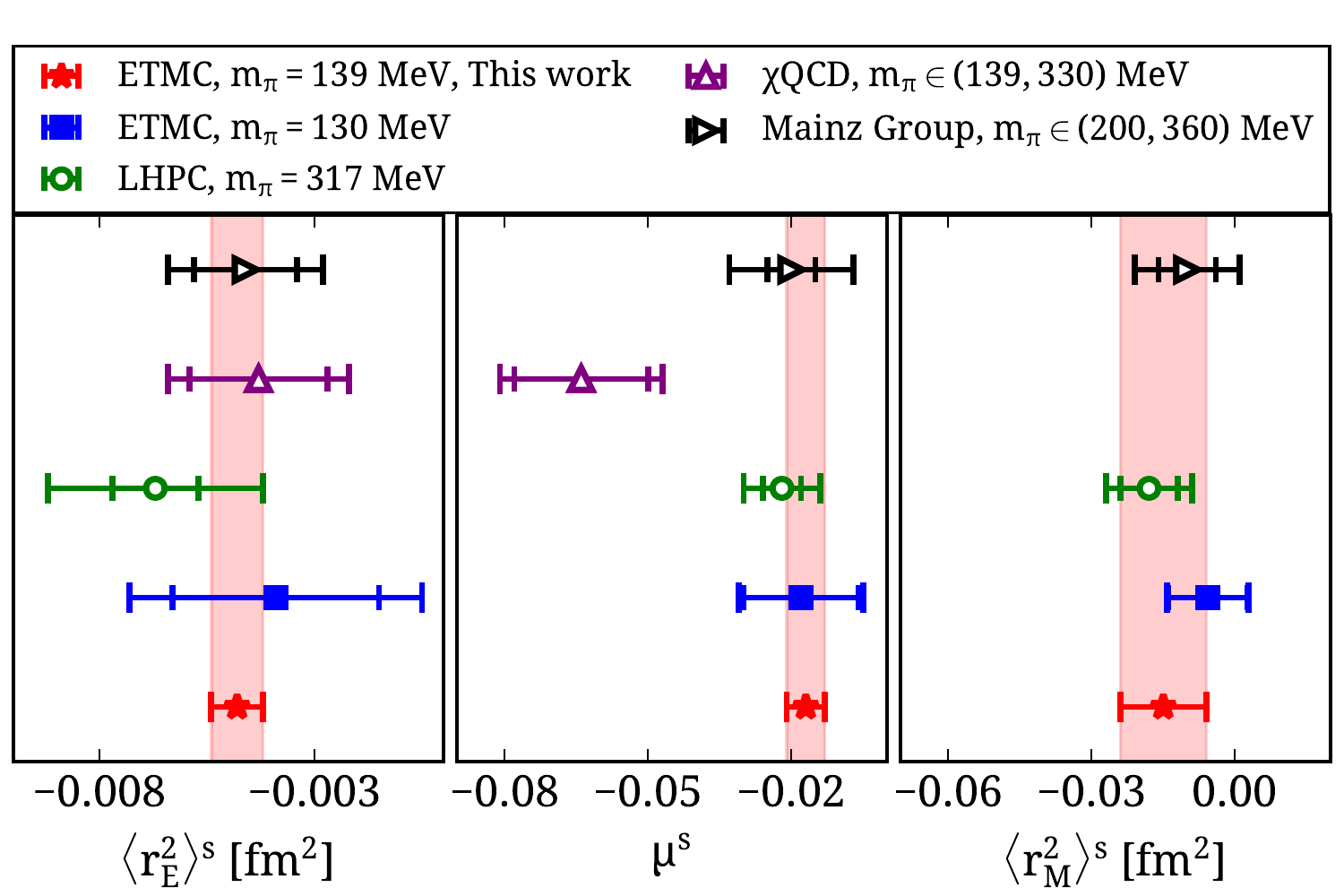}  \\
  \caption{ The left most panel shows results  for $\<r^2_E\>^s$, the middle panel for 
  $\mu^s$ and  the right panel for $\<r^2_M\>^s$. Results extracted using the $N_f=2+1+1$ cB211.072.64 ensemble are shown with the red stars and accompanied red error band. Results using the $N_f=2$ cA2.09.48 twisted mass ensemble are shown by the blue filled square~\cite{Alexandrou:2018zdf}.  
  We denote with open symbols results that include ensembles with larger than physical pion masses to  extract the value at the physical point.
  Results from  the
  $\chi$QCD~\cite{Sufian:2016pex} collaboration (purple upper triangles), Ref.~\cite{Djukanovic:2019jtp} (black right triangles) and from the LHPC ~\cite{Green:2015wqa}(green circles). The inner error bars indicate the statistical while the outer the total which includes systematic errors.} 
  \label{Fig:Comp}
\end{figure}

In  Fig.~\ref{Fig:Comp},  we show a comparison of the magnetic moment and radii  using the two twisted mass ensembles with the corresponding results from  the aforementioned groups.  As can be seen, there is an overall agreement. Our very precise values for the electric radius and magnetic moment clearly  confirm a non-zero value for both. The agreement among lattice QCD results using ensembles of different values of the  lattice spacings and volumes also indicates that cut-off and finite volume effects are small. This allows us to make a
comparison of  our results obtained using the $N_f=2$ and the $N_f=2+1+1$  ensembles to check for unquenching effects of the strange quark. The current statistical accuracy reveals no such effects.

\begin{figure}[ht!]
  \includegraphics[scale=0.45]{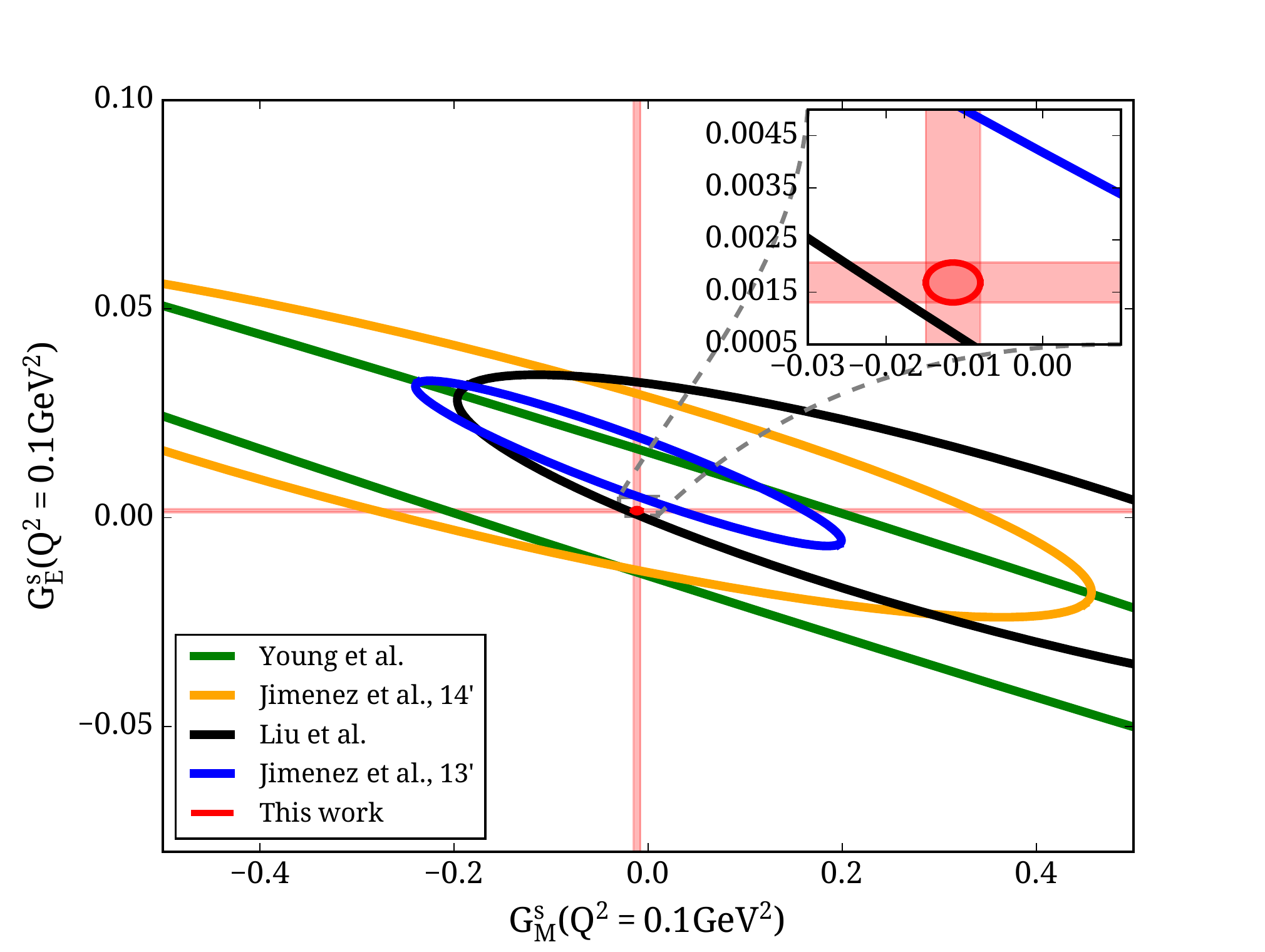}
  \caption{The red bands show the constrains arising from the values of $G_E^s$ and $G_M^s$ at $Q^2=0.1$~GeV extracted in this work. The ellipses indicate 95\% confidence level. The green ellipse is from Ref.~\cite{Young:2006jc}, the orange from Ref.~\cite{Gonzalez-Jimenez:2014bia}, the black from Ref.~\cite{Liu:2007yi} and the blue from Ref.~\cite{GonzalezJimenez:2011fq}.}
  \label{fig:constrains}
  \end{figure}

{\it{Conclusions:}}
A high precision calculation of the
strange nucleon electromagnetic form factors is obtained  using ensembles simulated with physical pion mass. Using the model independent z-expansion to fit the form factors we
obtain the following values for the radii and magnetic moment
\beq
\<r^2_E\>^s & =& -0.0048(6)\,{\rm fm}^2, \nonumber\\
\<r^2_M\>^s &= &-0.015(9)\,{\rm fm}^2, \nonumber \\
\mu^s &=& -0.017(4),
\label{final}
\eeq
clearly excluding a zero value for all three quantities.
This is a significant finding given the status of experimental searches where the results are inconclusive. 
For example the SAMPLE experimental data~\cite{Spayde:2003nr}  finds a strange magnetic moment of  $\mu^s= 0.37 \pm 0.20 \pm 0.26 \pm 0.15$  that is positive but also compatible with zero. 
More recently, the HAPPEX collaboration finds at $Q^2 \sim 0.62$~GeV$^2$~\cite{Ahmed:2011vp}  a negative value for $G_M^s = -0.070 \pm 0.067$, which again does not exclude zero. The G0 collaboration
 reported an upper bound of 10\% on the strange quark contributions as compared to  the total nucleon electromagnetic form factors~\cite{Androic:2009aa}.
  The A4 experiment~\cite{Baunack:2009gy},  also reported results consistent with zero strangeness, namely
 $G_E^s = 0.050 \pm 0.038 \pm 0.019$ and $G_M^s = −0.14 \pm 0.11 \pm 0.11$ at $Q^2=0.22$~GeV$^2$.
The Q-weak experiment~\cite{Zhao:2017xej,Benesch:2014bas,Armstrong:2012ps} is aiming to measure the weak charge of the proton to unprecedented accuracy to set limits on new physics. The strange electromagnetic form factors are a crucial input that can aid the interpretation of the experimental results.
In addition, the MESA~\cite{Becker:2018ggl} facility at Mainz targets very low $Q^2$ in order to improve the determination of the strange electric form factor, making these results of high relevance.
In Fig.~\ref{fig:constrains} we show the impact of the determination of the strange form factors on the experimental measurements at a given value of $Q^2=0.1$~GeV$^2$. Our values provide a stringent constrain on experimental searches.

\vspace*{0.3cm}
\emph{Acknowledgements: }
We would like to thank all members of the Extended Twisted Mass Collaboration (ETMC) for a very constructive and enjoyable collaboration.
M.C. acknowledges financial support by the U.S. National Science Foundation under Grant No.\ PHY-1714407.
This project has received funding from the Horizon 2020 research and innovation program
of the European Commission under the Marie Sk\l{}odowska-Curie grant agreement No 642069.
S.B. is supported by  the Marie Sk\l{}odowska-Curie grant agreement No 642069 of the European commission and from the project  COMPLEMENTARY/0916/0015 funded by the Cyprus Research Promotion Foundation.
The project used resources of the  SuperMUC
at Leibniz Supercomputing Centre under the Gauss Centre for Supercomputing e.V.  project pr74yo and of  Piz Daint at the Centro Svizzero di Calcolo Scientifico 
under the project  s702.
It also used XSEDE computational resources 
 supported by National Science Foundation grant number TG-PHY170022 as well as  on the Jureca system of the research center in J\"ulich, under the project ECY00.

\bibliography{refs}

\begin{thebibliography}{68}
\expandafter\ifx\csname natexlab\endcsname\relax\def\natexlab#1{#1}\fi
\expandafter\ifx\csname bibnamefont\endcsname\relax
  \def\bibnamefont#1{#1}\fi
\expandafter\ifx\csname bibfnamefont\endcsname\relax
  \def\bibfnamefont#1{#1}\fi
\expandafter\ifx\csname citenamefont\endcsname\relax
  \def\citenamefont#1{#1}\fi
\expandafter\ifx\csname url\endcsname\relax
  \def\url#1{\texttt{#1}}\fi
\expandafter\ifx\csname urlprefix\endcsname\relax\def\urlprefix{URL }\fi
\providecommand{\bibinfo}[2]{#2}
\providecommand{\eprint}[2][]{\url{#2}}

\bibitem[{\citenamefont{Spayde et~al.}(2004)}]{Spayde:2003nr}
\bibinfo{author}{\bibfnamefont{D.~T.} \bibnamefont{Spayde}}
  \bibnamefont{et~al.} (\bibinfo{collaboration}{SAMPLE}),
  \bibinfo{journal}{Phys. Lett.} \textbf{\bibinfo{volume}{B583}},
  \bibinfo{pages}{79} (\bibinfo{year}{2004}), \eprint{nucl-ex/0312016}.

\bibitem[{\citenamefont{Beise et~al.}(2005)\citenamefont{Beise, Pitt, and
  Spayde}}]{Beise:2004py}
\bibinfo{author}{\bibfnamefont{E.~J.} \bibnamefont{Beise}},
  \bibinfo{author}{\bibfnamefont{M.~L.} \bibnamefont{Pitt}}, \bibnamefont{and}
  \bibinfo{author}{\bibfnamefont{D.~T.} \bibnamefont{Spayde}},
  \bibinfo{journal}{Prog. Part. Nucl. Phys.} \textbf{\bibinfo{volume}{54}},
  \bibinfo{pages}{289} (\bibinfo{year}{2005}), \eprint{nucl-ex/0412054}.

\bibitem[{\citenamefont{Maas et~al.}(2004)}]{Maas:2004ta}
\bibinfo{author}{\bibfnamefont{F.~E.} \bibnamefont{Maas}} \bibnamefont{et~al.}
  (\bibinfo{collaboration}{A4}), \bibinfo{journal}{Phys. Rev. Lett.}
  \textbf{\bibinfo{volume}{93}}, \bibinfo{pages}{022002}
  (\bibinfo{year}{2004}), \eprint{nucl-ex/0401019}.

\bibitem[{\citenamefont{Maas et~al.}(2005)}]{Maas:2004dh}
\bibinfo{author}{\bibfnamefont{F.~E.} \bibnamefont{Maas}} \bibnamefont{et~al.},
  \bibinfo{journal}{Phys. Rev. Lett.} \textbf{\bibinfo{volume}{94}},
  \bibinfo{pages}{152001} (\bibinfo{year}{2005}), \eprint{nucl-ex/0412030}.

\bibitem[{\citenamefont{Baunack et~al.}(2009)}]{Baunack:2009gy}
\bibinfo{author}{\bibfnamefont{S.}~\bibnamefont{Baunack}} \bibnamefont{et~al.},
  \bibinfo{journal}{Phys. Rev. Lett.} \textbf{\bibinfo{volume}{102}},
  \bibinfo{pages}{151803} (\bibinfo{year}{2009}), \eprint{0903.2733}.

\bibitem[{\citenamefont{Aniol et~al.}(2006{\natexlab{a}})}]{Aniol:2005zf}
\bibinfo{author}{\bibfnamefont{K.~A.} \bibnamefont{Aniol}} \bibnamefont{et~al.}
  (\bibinfo{collaboration}{HAPPEX}), \bibinfo{journal}{Phys. Rev. Lett.}
  \textbf{\bibinfo{volume}{96}}, \bibinfo{pages}{022003}
  (\bibinfo{year}{2006}{\natexlab{a}}), \eprint{nucl-ex/0506010}.

\bibitem[{\citenamefont{Aniol et~al.}(2006{\natexlab{b}})}]{Aniol:2005zg}
\bibinfo{author}{\bibfnamefont{K.~A.} \bibnamefont{Aniol}} \bibnamefont{et~al.}
  (\bibinfo{collaboration}{HAPPEX}), \bibinfo{journal}{Phys. Lett.}
  \textbf{\bibinfo{volume}{B635}}, \bibinfo{pages}{275}
  (\bibinfo{year}{2006}{\natexlab{b}}), \eprint{nucl-ex/0506011}.

\bibitem[{\citenamefont{Acha et~al.}(2007)}]{Acha:2006my}
\bibinfo{author}{\bibfnamefont{A.}~\bibnamefont{Acha}} \bibnamefont{et~al.}
  (\bibinfo{collaboration}{HAPPEX}), \bibinfo{journal}{Phys. Rev. Lett.}
  \textbf{\bibinfo{volume}{98}}, \bibinfo{pages}{032301}
  (\bibinfo{year}{2007}), \eprint{nucl-ex/0609002}.

\bibitem[{\citenamefont{Ahmed et~al.}(2012)}]{Ahmed:2011vp}
\bibinfo{author}{\bibfnamefont{Z.}~\bibnamefont{Ahmed}} \bibnamefont{et~al.}
  (\bibinfo{collaboration}{HAPPEX}), \bibinfo{journal}{Phys. Rev. Lett.}
  \textbf{\bibinfo{volume}{108}}, \bibinfo{pages}{102001}
  (\bibinfo{year}{2012}), \eprint{1107.0913}.

\bibitem[{\citenamefont{Armstrong et~al.}(2005)}]{Armstrong:2005hs}
\bibinfo{author}{\bibfnamefont{D.~S.} \bibnamefont{Armstrong}}
  \bibnamefont{et~al.} (\bibinfo{collaboration}{G0}), \bibinfo{journal}{Phys.
  Rev. Lett.} \textbf{\bibinfo{volume}{95}}, \bibinfo{pages}{092001}
  (\bibinfo{year}{2005}), \eprint{nucl-ex/0506021}.

\bibitem[{\citenamefont{Androic et~al.}(2010)}]{Androic:2009aa}
\bibinfo{author}{\bibfnamefont{D.}~\bibnamefont{Androic}} \bibnamefont{et~al.}
  (\bibinfo{collaboration}{G0}), \bibinfo{journal}{Phys. Rev. Lett.}
  \textbf{\bibinfo{volume}{104}}, \bibinfo{pages}{012001}
  (\bibinfo{year}{2010}), \eprint{0909.5107}.

\bibitem[{\citenamefont{González-Jiménez
  et~al.}(2014)\citenamefont{González-Jiménez, Caballero, and
  Donnelly}}]{Gonzalez-Jimenez:2014bia}
\bibinfo{author}{\bibfnamefont{R.}~\bibnamefont{González-Jiménez}},
  \bibinfo{author}{\bibfnamefont{J.~A.} \bibnamefont{Caballero}},
  \bibnamefont{and} \bibinfo{author}{\bibfnamefont{T.~W.}
  \bibnamefont{Donnelly}}, \bibinfo{journal}{Phys. Rev.}
  \textbf{\bibinfo{volume}{D90}}, \bibinfo{pages}{033002}
  (\bibinfo{year}{2014}), \eprint{1403.5119}.

\bibitem[{\citenamefont{Maas and Paschke}(2017)}]{Maas:2017snj}
\bibinfo{author}{\bibfnamefont{F.~E.} \bibnamefont{Maas}} \bibnamefont{and}
  \bibinfo{author}{\bibfnamefont{K.~D.} \bibnamefont{Paschke}},
  \bibinfo{journal}{Prog. Part. Nucl. Phys.} \textbf{\bibinfo{volume}{95}},
  \bibinfo{pages}{209} (\bibinfo{year}{2017}).

\bibitem[{\citenamefont{Weigel et~al.}(1995)\citenamefont{Weigel, Abada,
  Alkofer, and Reinhardt}}]{Weigel:1995jc}
\bibinfo{author}{\bibfnamefont{H.}~\bibnamefont{Weigel}},
  \bibinfo{author}{\bibfnamefont{A.}~\bibnamefont{Abada}},
  \bibinfo{author}{\bibfnamefont{R.}~\bibnamefont{Alkofer}}, \bibnamefont{and}
  \bibinfo{author}{\bibfnamefont{H.}~\bibnamefont{Reinhardt}},
  \bibinfo{journal}{Phys. Lett.} \textbf{\bibinfo{volume}{B353}},
  \bibinfo{pages}{20} (\bibinfo{year}{1995}), \eprint{hep-ph/9503241}.

\bibitem[{\citenamefont{Lyubovitskij et~al.}(2002)\citenamefont{Lyubovitskij,
  Wang, Gutsche, and Faessler}}]{Lyubovitskij:2002ng}
\bibinfo{author}{\bibfnamefont{V.~E.} \bibnamefont{Lyubovitskij}},
  \bibinfo{author}{\bibfnamefont{P.}~\bibnamefont{Wang}},
  \bibinfo{author}{\bibfnamefont{T.}~\bibnamefont{Gutsche}}, \bibnamefont{and}
  \bibinfo{author}{\bibfnamefont{A.}~\bibnamefont{Faessler}},
  \bibinfo{journal}{Phys. Rev.} \textbf{\bibinfo{volume}{C66}},
  \bibinfo{pages}{055204} (\bibinfo{year}{2002}), \eprint{hep-ph/0207225}.

\bibitem[{\citenamefont{Silva et~al.}(2006)\citenamefont{Silva, Kim, Urbano,
  and Goeke}}]{Silva:2005qm}
\bibinfo{author}{\bibfnamefont{A.}~\bibnamefont{Silva}},
  \bibinfo{author}{\bibfnamefont{H.-C.} \bibnamefont{Kim}},
  \bibinfo{author}{\bibfnamefont{D.}~\bibnamefont{Urbano}}, \bibnamefont{and}
  \bibinfo{author}{\bibfnamefont{K.}~\bibnamefont{Goeke}},
  \bibinfo{journal}{Phys. Rev.} \textbf{\bibinfo{volume}{D74}},
  \bibinfo{pages}{054011} (\bibinfo{year}{2006}), \eprint{hep-ph/0601239}.

\bibitem[{\citenamefont{Goeke et~al.}(2007)\citenamefont{Goeke, Kim, Silva, and
  Urbano}}]{Goeke:2006gi}
\bibinfo{author}{\bibfnamefont{K.}~\bibnamefont{Goeke}},
  \bibinfo{author}{\bibfnamefont{H.-C.} \bibnamefont{Kim}},
  \bibinfo{author}{\bibfnamefont{A.}~\bibnamefont{Silva}}, \bibnamefont{and}
  \bibinfo{author}{\bibfnamefont{D.}~\bibnamefont{Urbano}},
  \bibinfo{journal}{Eur. Phys. J.} \textbf{\bibinfo{volume}{A32}},
  \bibinfo{pages}{393} (\bibinfo{year}{2007}), \bibinfo{note}{[,23(2006)]},
  \eprint{hep-ph/0608262}.

\bibitem[{\citenamefont{Bijker}(2006)}]{Bijker:2005pe}
\bibinfo{author}{\bibfnamefont{R.}~\bibnamefont{Bijker}}, \bibinfo{journal}{J.
  Phys.} \textbf{\bibinfo{volume}{G32}}, \bibinfo{pages}{L49}
  (\bibinfo{year}{2006}), \eprint{nucl-th/0511060}.

\bibitem[{\citenamefont{Wang et~al.}(2014)\citenamefont{Wang, Leinweber, and
  Thomas}}]{Wang:2013cfp}
\bibinfo{author}{\bibfnamefont{P.}~\bibnamefont{Wang}},
  \bibinfo{author}{\bibfnamefont{D.~B.} \bibnamefont{Leinweber}},
  \bibnamefont{and} \bibinfo{author}{\bibfnamefont{A.~W.}
  \bibnamefont{Thomas}}, \bibinfo{journal}{Phys. Rev.}
  \textbf{\bibinfo{volume}{D89}}, \bibinfo{pages}{033008}
  (\bibinfo{year}{2014}), \eprint{1312.3375}.

\bibitem[{\citenamefont{Hobbs et~al.}(2015)\citenamefont{Hobbs, Alberg, and
  Miller}}]{Hobbs:2014lea}
\bibinfo{author}{\bibfnamefont{T.~J.} \bibnamefont{Hobbs}},
  \bibinfo{author}{\bibfnamefont{M.}~\bibnamefont{Alberg}}, \bibnamefont{and}
  \bibinfo{author}{\bibfnamefont{G.~A.} \bibnamefont{Miller}},
  \bibinfo{journal}{Phys. Rev.} \textbf{\bibinfo{volume}{C91}},
  \bibinfo{pages}{035205} (\bibinfo{year}{2015}), \eprint{1412.4871}.

\bibitem[{\citenamefont{Green et~al.}(2015)\citenamefont{Green, Meinel,
  Engelhardt, Krieg, Laeuchli, Negele, Orginos, Pochinsky, and
  Syritsyn}}]{Green:2015wqa}
\bibinfo{author}{\bibfnamefont{J.}~\bibnamefont{Green}},
  \bibinfo{author}{\bibfnamefont{S.}~\bibnamefont{Meinel}},
  \bibinfo{author}{\bibfnamefont{M.}~\bibnamefont{Engelhardt}},
  \bibinfo{author}{\bibfnamefont{S.}~\bibnamefont{Krieg}},
  \bibinfo{author}{\bibfnamefont{J.}~\bibnamefont{Laeuchli}},
  \bibinfo{author}{\bibfnamefont{J.}~\bibnamefont{Negele}},
  \bibinfo{author}{\bibfnamefont{K.}~\bibnamefont{Orginos}},
  \bibinfo{author}{\bibfnamefont{A.}~\bibnamefont{Pochinsky}},
  \bibnamefont{and} \bibinfo{author}{\bibfnamefont{S.}~\bibnamefont{Syritsyn}},
  \bibinfo{journal}{Phys. Rev.} \textbf{\bibinfo{volume}{D92}},
  \bibinfo{pages}{031501} (\bibinfo{year}{2015}), \eprint{1505.01803}.

\bibitem[{\citenamefont{Sufian et~al.}(2017{\natexlab{a}})\citenamefont{Sufian,
  Yang, Alexandru, Draper, Liang, and Liu}}]{Sufian:2016pex}
\bibinfo{author}{\bibfnamefont{R.~S.} \bibnamefont{Sufian}},
  \bibinfo{author}{\bibfnamefont{Y.-B.} \bibnamefont{Yang}},
  \bibinfo{author}{\bibfnamefont{A.}~\bibnamefont{Alexandru}},
  \bibinfo{author}{\bibfnamefont{T.}~\bibnamefont{Draper}},
  \bibinfo{author}{\bibfnamefont{J.}~\bibnamefont{Liang}}, \bibnamefont{and}
  \bibinfo{author}{\bibfnamefont{K.-F.} \bibnamefont{Liu}},
  \bibinfo{journal}{Phys. Rev. Lett.} \textbf{\bibinfo{volume}{118}},
  \bibinfo{pages}{042001} (\bibinfo{year}{2017}{\natexlab{a}}),
  \eprint{1606.07075}.

\bibitem[{\citenamefont{Sufian et~al.}(2017{\natexlab{b}})\citenamefont{Sufian,
  Yang, Liang, Draper, and Liu}}]{Sufian:2017osl}
\bibinfo{author}{\bibfnamefont{R.~S.} \bibnamefont{Sufian}},
  \bibinfo{author}{\bibfnamefont{Y.-B.} \bibnamefont{Yang}},
  \bibinfo{author}{\bibfnamefont{J.}~\bibnamefont{Liang}},
  \bibinfo{author}{\bibfnamefont{T.}~\bibnamefont{Draper}}, \bibnamefont{and}
  \bibinfo{author}{\bibfnamefont{K.-F.} \bibnamefont{Liu}},
  \bibinfo{journal}{Phys. Rev.} \textbf{\bibinfo{volume}{D96}},
  \bibinfo{pages}{114504} (\bibinfo{year}{2017}{\natexlab{b}}),
  \eprint{1705.05849}.

\bibitem[{\citenamefont{Alexandrou
  et~al.}(2018{\natexlab{a}})\citenamefont{Alexandrou, Constantinou,
  Hadjiyiannakou, Jansen, Kallidonis, Koutsou, and Vaquero
  Avil\'es-Casco}}]{Alexandrou:2018zdf}
\bibinfo{author}{\bibfnamefont{C.}~\bibnamefont{Alexandrou}},
  \bibinfo{author}{\bibfnamefont{M.}~\bibnamefont{Constantinou}},
  \bibinfo{author}{\bibfnamefont{K.}~\bibnamefont{Hadjiyiannakou}},
  \bibinfo{author}{\bibfnamefont{K.}~\bibnamefont{Jansen}},
  \bibinfo{author}{\bibfnamefont{C.}~\bibnamefont{Kallidonis}},
  \bibinfo{author}{\bibfnamefont{G.}~\bibnamefont{Koutsou}}, \bibnamefont{and}
  \bibinfo{author}{\bibfnamefont{A.}~\bibnamefont{Vaquero Avil\'es-Casco}},
  \bibinfo{journal}{Phys. Rev.} \textbf{\bibinfo{volume}{D97}},
  \bibinfo{pages}{094504} (\bibinfo{year}{2018}{\natexlab{a}}),
  \eprint{1801.09581}.

\bibitem[{\citenamefont{Alexandrou
  et~al.}(2018{\natexlab{b}})}]{Alexandrou:2018egz}
\bibinfo{author}{\bibfnamefont{C.}~\bibnamefont{Alexandrou}}
  \bibnamefont{et~al.}, \bibinfo{journal}{Phys. Rev.}
  \textbf{\bibinfo{volume}{D98}}, \bibinfo{pages}{054518}
  (\bibinfo{year}{2018}{\natexlab{b}}), \eprint{1807.00495}.

\bibitem[{\citenamefont{Frezzotti et~al.}(2001)\citenamefont{Frezzotti, Grassi,
  Sint, and Weisz}}]{Frezzotti:2000nk}
\bibinfo{author}{\bibfnamefont{R.}~\bibnamefont{Frezzotti}},
  \bibinfo{author}{\bibfnamefont{P.~A.} \bibnamefont{Grassi}},
  \bibinfo{author}{\bibfnamefont{S.}~\bibnamefont{Sint}}, \bibnamefont{and}
  \bibinfo{author}{\bibfnamefont{P.}~\bibnamefont{Weisz}}
  (\bibinfo{collaboration}{Alpha}), \bibinfo{journal}{JHEP}
  \textbf{\bibinfo{volume}{08}}, \bibinfo{pages}{058} (\bibinfo{year}{2001}),
  \eprint{hep-lat/0101001}.

\bibitem[{\citenamefont{Frezzotti and
  Rossi}(2004{\natexlab{a}})}]{Frezzotti:2003ni}
\bibinfo{author}{\bibfnamefont{R.}~\bibnamefont{Frezzotti}} \bibnamefont{and}
  \bibinfo{author}{\bibfnamefont{G.~C.} \bibnamefont{Rossi}},
  \bibinfo{journal}{JHEP} \textbf{\bibinfo{volume}{08}}, \bibinfo{pages}{007}
  (\bibinfo{year}{2004}{\natexlab{a}}), \eprint{hep-lat/0306014}.

\bibitem[{\citenamefont{Frezzotti and
  Rossi}(2004{\natexlab{b}})}]{Frezzotti:2004wz}
\bibinfo{author}{\bibfnamefont{R.}~\bibnamefont{Frezzotti}} \bibnamefont{and}
  \bibinfo{author}{\bibfnamefont{G.~C.} \bibnamefont{Rossi}},
  \bibinfo{journal}{JHEP} \textbf{\bibinfo{volume}{10}}, \bibinfo{pages}{070}
  (\bibinfo{year}{2004}{\natexlab{b}}), \eprint{hep-lat/0407002}.

\bibitem[{\citenamefont{Sheikholeslami and
  Wohlert}(1985)}]{Sheikholeslami:1985ij}
\bibinfo{author}{\bibfnamefont{B.}~\bibnamefont{Sheikholeslami}}
  \bibnamefont{and} \bibinfo{author}{\bibfnamefont{R.}~\bibnamefont{Wohlert}},
  \bibinfo{journal}{Nucl. Phys.} \textbf{\bibinfo{volume}{B259}},
  \bibinfo{pages}{572} (\bibinfo{year}{1985}).

\bibitem[{\citenamefont{Alexandrou
  et~al.}(2018{\natexlab{c}})\citenamefont{Alexandrou, Bacchio, Constantinou,
  Finkenrath, Hadjiyiannakou, Jansen, Koutsou, and Casco}}]{Alexandrou:2018sjm}
\bibinfo{author}{\bibfnamefont{C.}~\bibnamefont{Alexandrou}},
  \bibinfo{author}{\bibfnamefont{S.}~\bibnamefont{Bacchio}},
  \bibinfo{author}{\bibfnamefont{M.}~\bibnamefont{Constantinou}},
  \bibinfo{author}{\bibfnamefont{J.}~\bibnamefont{Finkenrath}},
  \bibinfo{author}{\bibfnamefont{K.}~\bibnamefont{Hadjiyiannakou}},
  \bibinfo{author}{\bibfnamefont{K.}~\bibnamefont{Jansen}},
  \bibinfo{author}{\bibfnamefont{G.}~\bibnamefont{Koutsou}}, \bibnamefont{and}
  \bibinfo{author}{\bibfnamefont{A.~V.~A.} \bibnamefont{Casco}}
  (\bibinfo{year}{2018}{\natexlab{c}}), \eprint{1812.10311}.

\bibitem[{\citenamefont{Alexandrou et~al.}(1994)\citenamefont{Alexandrou,
  Gusken, Jegerlehner, Schilling, and Sommer}}]{Alexandrou:1992ti}
\bibinfo{author}{\bibfnamefont{C.}~\bibnamefont{Alexandrou}},
  \bibinfo{author}{\bibfnamefont{S.}~\bibnamefont{Gusken}},
  \bibinfo{author}{\bibfnamefont{F.}~\bibnamefont{Jegerlehner}},
  \bibinfo{author}{\bibfnamefont{K.}~\bibnamefont{Schilling}},
  \bibnamefont{and} \bibinfo{author}{\bibfnamefont{R.}~\bibnamefont{Sommer}},
  \bibinfo{journal}{Nucl. Phys.} \textbf{\bibinfo{volume}{B414}},
  \bibinfo{pages}{815} (\bibinfo{year}{1994}), \eprint{hep-lat/9211042}.

\bibitem[{\citenamefont{Gusken}(1990)}]{Gusken:1989qx}
\bibinfo{author}{\bibfnamefont{S.}~\bibnamefont{Gusken}},
  \bibinfo{journal}{Nucl. Phys. Proc. Suppl.} \textbf{\bibinfo{volume}{17}},
  \bibinfo{pages}{361} (\bibinfo{year}{1990}).

\bibitem[{\citenamefont{Albanese et~al.}(1987)}]{Albanese:1987ds}
\bibinfo{author}{\bibfnamefont{M.}~\bibnamefont{Albanese}} \bibnamefont{et~al.}
  (\bibinfo{collaboration}{APE}), \bibinfo{journal}{Phys. Lett.}
  \textbf{\bibinfo{volume}{B192}}, \bibinfo{pages}{163} (\bibinfo{year}{1987}).

\bibitem[{\citenamefont{Wilcox}(1999)}]{Wilcox:1999ab}
\bibinfo{author}{\bibfnamefont{W.}~\bibnamefont{Wilcox}}, in
  \emph{\bibinfo{booktitle}{{Numerical challenges in lattice quantum
  chromodynamics. Proceedings, Joint Interdisciplinary Workshop, Wuppertal,
  Germany, August 22-24, 1999}}} (\bibinfo{year}{1999}), pp.
  \bibinfo{pages}{127--141}, \eprint{hep-lat/9911013}.

\bibitem[{\citenamefont{Foley et~al.}(2005)\citenamefont{Foley, Jimmy~Juge,
  O'Cais, Peardon, Ryan, and Skullerud}}]{Foley:2005ac}
\bibinfo{author}{\bibfnamefont{J.}~\bibnamefont{Foley}},
  \bibinfo{author}{\bibfnamefont{K.}~\bibnamefont{Jimmy~Juge}},
  \bibinfo{author}{\bibfnamefont{A.}~\bibnamefont{O'Cais}},
  \bibinfo{author}{\bibfnamefont{M.}~\bibnamefont{Peardon}},
  \bibinfo{author}{\bibfnamefont{S.~M.} \bibnamefont{Ryan}}, \bibnamefont{and}
  \bibinfo{author}{\bibfnamefont{J.-I.} \bibnamefont{Skullerud}},
  \bibinfo{journal}{Comput. Phys. Commun.} \textbf{\bibinfo{volume}{172}},
  \bibinfo{pages}{145} (\bibinfo{year}{2005}), \eprint{hep-lat/0505023}.

\bibitem[{\citenamefont{Alexandrou
  et~al.}(2019{\natexlab{a}})}]{Alexandrou:2019ali}
\bibinfo{author}{\bibfnamefont{C.}~\bibnamefont{Alexandrou}}
  \bibnamefont{et~al.} (\bibinfo{year}{2019}{\natexlab{a}}),
  \eprint{1908.10706}.

\bibitem[{\citenamefont{Alexandrou
  et~al.}(2012{\natexlab{a}})\citenamefont{Alexandrou, Hadjiyiannakou, Koutsou,
  O'Cais, and Strelchenko}}]{Alexandrou:2012zz}
\bibinfo{author}{\bibfnamefont{C.}~\bibnamefont{Alexandrou}},
  \bibinfo{author}{\bibfnamefont{K.}~\bibnamefont{Hadjiyiannakou}},
  \bibinfo{author}{\bibfnamefont{G.}~\bibnamefont{Koutsou}},
  \bibinfo{author}{\bibfnamefont{A.}~\bibnamefont{O'Cais}}, \bibnamefont{and}
  \bibinfo{author}{\bibfnamefont{A.}~\bibnamefont{Strelchenko}},
  \bibinfo{journal}{Comput. Phys. Commun.} \textbf{\bibinfo{volume}{183}},
  \bibinfo{pages}{1215} (\bibinfo{year}{2012}{\natexlab{a}}),
  \eprint{1108.2473}.

\bibitem[{\citenamefont{Stathopoulos et~al.}(2013)\citenamefont{Stathopoulos,
  Laeuchli, and Orginos}}]{Stathopoulos:2013aci}
\bibinfo{author}{\bibfnamefont{A.}~\bibnamefont{Stathopoulos}},
  \bibinfo{author}{\bibfnamefont{J.}~\bibnamefont{Laeuchli}}, \bibnamefont{and}
  \bibinfo{author}{\bibfnamefont{K.}~\bibnamefont{Orginos}}
  (\bibinfo{year}{2013}), \eprint{1302.4018}.

\bibitem[{\citenamefont{Michael and Urbach}(2007)}]{Michael:2007vn}
\bibinfo{author}{\bibfnamefont{C.}~\bibnamefont{Michael}} \bibnamefont{and}
  \bibinfo{author}{\bibfnamefont{C.}~\bibnamefont{Urbach}}
  (\bibinfo{collaboration}{ETM}), \bibinfo{journal}{PoS}
  \textbf{\bibinfo{volume}{LATTICE2007}}, \bibinfo{pages}{122}
  (\bibinfo{year}{2007}), \eprint{0709.4564}.

\bibitem[{\citenamefont{McNeile and Michael}(2006)}]{McNeile:2006bz}
\bibinfo{author}{\bibfnamefont{C.}~\bibnamefont{McNeile}} \bibnamefont{and}
  \bibinfo{author}{\bibfnamefont{C.}~\bibnamefont{Michael}}
  (\bibinfo{collaboration}{UKQCD}), \bibinfo{journal}{Phys. Rev.}
  \textbf{\bibinfo{volume}{D73}}, \bibinfo{pages}{074506}
  (\bibinfo{year}{2006}), \eprint{hep-lat/0603007}.

\bibitem[{\citenamefont{Alexandrou
  et~al.}(2014{\natexlab{a}})\citenamefont{Alexandrou, Constantinou, Drach,
  Hadjiyiannakou, Jansen, Koutsou, Strelchenko, and
  Vaquero}}]{Alexandrou:2013wca}
\bibinfo{author}{\bibfnamefont{C.}~\bibnamefont{Alexandrou}},
  \bibinfo{author}{\bibfnamefont{M.}~\bibnamefont{Constantinou}},
  \bibinfo{author}{\bibfnamefont{V.}~\bibnamefont{Drach}},
  \bibinfo{author}{\bibfnamefont{K.}~\bibnamefont{Hadjiyiannakou}},
  \bibinfo{author}{\bibfnamefont{K.}~\bibnamefont{Jansen}},
  \bibinfo{author}{\bibfnamefont{G.}~\bibnamefont{Koutsou}},
  \bibinfo{author}{\bibfnamefont{A.}~\bibnamefont{Strelchenko}},
  \bibnamefont{and} \bibinfo{author}{\bibfnamefont{A.}~\bibnamefont{Vaquero}},
  \bibinfo{journal}{Comput. Phys. Commun.} \textbf{\bibinfo{volume}{185}},
  \bibinfo{pages}{1370} (\bibinfo{year}{2014}{\natexlab{a}}),
  \eprint{1309.2256}.

\bibitem[{\citenamefont{Abdel-Rehim et~al.}(2014)\citenamefont{Abdel-Rehim,
  Alexandrou, Constantinou, Drach, Hadjiyiannakou, Jansen, Koutsou, and
  Vaquero}}]{Abdel-Rehim:2013wlz}
\bibinfo{author}{\bibfnamefont{A.}~\bibnamefont{Abdel-Rehim}},
  \bibinfo{author}{\bibfnamefont{C.}~\bibnamefont{Alexandrou}},
  \bibinfo{author}{\bibfnamefont{M.}~\bibnamefont{Constantinou}},
  \bibinfo{author}{\bibfnamefont{V.}~\bibnamefont{Drach}},
  \bibinfo{author}{\bibfnamefont{K.}~\bibnamefont{Hadjiyiannakou}},
  \bibinfo{author}{\bibfnamefont{K.}~\bibnamefont{Jansen}},
  \bibinfo{author}{\bibfnamefont{G.}~\bibnamefont{Koutsou}}, \bibnamefont{and}
  \bibinfo{author}{\bibfnamefont{A.}~\bibnamefont{Vaquero}},
  \bibinfo{journal}{Phys. Rev.} \textbf{\bibinfo{volume}{D89}},
  \bibinfo{pages}{034501} (\bibinfo{year}{2014}), \eprint{1310.6339}.

\bibitem[{\citenamefont{Babich et~al.}(2010)\citenamefont{Babich, Clark, and
  Joo}}]{Babich:2010mu}
\bibinfo{author}{\bibfnamefont{R.}~\bibnamefont{Babich}},
  \bibinfo{author}{\bibfnamefont{M.~A.} \bibnamefont{Clark}}, \bibnamefont{and}
  \bibinfo{author}{\bibfnamefont{B.}~\bibnamefont{Joo}}, in
  \emph{\bibinfo{booktitle}{{SC 10 (Supercomputing 2010) New Orleans,
  Louisiana, November 13-19, 2010}}} (\bibinfo{year}{2010}),
  \eprint{1011.0024},
  \urlprefix\url{http://www1.jlab.org/Ul/publications/view_pub.cfm?pub_id=10186}.

\bibitem[{\citenamefont{Clark et~al.}(2016)\citenamefont{Clark, Joó,
  Strelchenko, Cheng, Gambhir, and Brower}}]{Clark:2016rdz}
\bibinfo{author}{\bibfnamefont{M.~A.} \bibnamefont{Clark}},
  \bibinfo{author}{\bibfnamefont{B.}~\bibnamefont{Joó}},
  \bibinfo{author}{\bibfnamefont{A.}~\bibnamefont{Strelchenko}},
  \bibinfo{author}{\bibfnamefont{M.}~\bibnamefont{Cheng}},
  \bibinfo{author}{\bibfnamefont{A.}~\bibnamefont{Gambhir}}, \bibnamefont{and}
  \bibinfo{author}{\bibfnamefont{R.}~\bibnamefont{Brower}}
  (\bibinfo{year}{2016}), \eprint{1612.07873}.

\bibitem[{\citenamefont{Alexandrou
  et~al.}(2014{\natexlab{b}})\citenamefont{Alexandrou, Hadjiyiannakou, Koutsou,
  Strelchenko, and Avilés-Casco}}]{Alexandrou:2014fva}
\bibinfo{author}{\bibfnamefont{C.}~\bibnamefont{Alexandrou}},
  \bibinfo{author}{\bibfnamefont{K.}~\bibnamefont{Hadjiyiannakou}},
  \bibinfo{author}{\bibfnamefont{G.}~\bibnamefont{Koutsou}},
  \bibinfo{author}{\bibfnamefont{A.}~\bibnamefont{Strelchenko}},
  \bibnamefont{and} \bibinfo{author}{\bibfnamefont{A.~V.}
  \bibnamefont{Avilés-Casco}}, \bibinfo{journal}{PoS}
  \textbf{\bibinfo{volume}{LATTICE2013}}, \bibinfo{pages}{411}
  (\bibinfo{year}{2014}{\natexlab{b}}), \eprint{1401.6750}.

\bibitem[{\citenamefont{Alexandrou et~al.}(2013)\citenamefont{Alexandrou,
  Constantinou, Dinter, Drach, Jansen, Kallidonis, and
  Koutsou}}]{Alexandrou:2013joa}
\bibinfo{author}{\bibfnamefont{C.}~\bibnamefont{Alexandrou}},
  \bibinfo{author}{\bibfnamefont{M.}~\bibnamefont{Constantinou}},
  \bibinfo{author}{\bibfnamefont{S.}~\bibnamefont{Dinter}},
  \bibinfo{author}{\bibfnamefont{V.}~\bibnamefont{Drach}},
  \bibinfo{author}{\bibfnamefont{K.}~\bibnamefont{Jansen}},
  \bibinfo{author}{\bibfnamefont{C.}~\bibnamefont{Kallidonis}},
  \bibnamefont{and} \bibinfo{author}{\bibfnamefont{G.}~\bibnamefont{Koutsou}},
  \bibinfo{journal}{Phys. Rev.} \textbf{\bibinfo{volume}{D88}},
  \bibinfo{pages}{014509} (\bibinfo{year}{2013}), \eprint{1303.5979}.

\bibitem[{\citenamefont{Alexandrou
  et~al.}(2011{\natexlab{a}})\citenamefont{Alexandrou, Brinet, Carbonell,
  Constantinou, Harraud, Guichon, Jansen, Korzec, and
  Papinutto}}]{Alexandrou:2011db}
\bibinfo{author}{\bibfnamefont{C.}~\bibnamefont{Alexandrou}},
  \bibinfo{author}{\bibfnamefont{M.}~\bibnamefont{Brinet}},
  \bibinfo{author}{\bibfnamefont{J.}~\bibnamefont{Carbonell}},
  \bibinfo{author}{\bibfnamefont{M.}~\bibnamefont{Constantinou}},
  \bibinfo{author}{\bibfnamefont{P.~A.} \bibnamefont{Harraud}},
  \bibinfo{author}{\bibfnamefont{P.}~\bibnamefont{Guichon}},
  \bibinfo{author}{\bibfnamefont{K.}~\bibnamefont{Jansen}},
  \bibinfo{author}{\bibfnamefont{T.}~\bibnamefont{Korzec}}, \bibnamefont{and}
  \bibinfo{author}{\bibfnamefont{M.}~\bibnamefont{Papinutto}},
  \bibinfo{journal}{Phys. Rev.} \textbf{\bibinfo{volume}{D83}},
  \bibinfo{pages}{094502} (\bibinfo{year}{2011}{\natexlab{a}}),
  \eprint{1102.2208}.

\bibitem[{\citenamefont{Alexandrou et~al.}(2006)\citenamefont{Alexandrou,
  Koutsou, Negele, and Tsapalis}}]{Alexandrou:2006ru}
\bibinfo{author}{\bibfnamefont{C.}~\bibnamefont{Alexandrou}},
  \bibinfo{author}{\bibfnamefont{G.}~\bibnamefont{Koutsou}},
  \bibinfo{author}{\bibfnamefont{J.~W.} \bibnamefont{Negele}},
  \bibnamefont{and} \bibinfo{author}{\bibfnamefont{A.}~\bibnamefont{Tsapalis}},
  \bibinfo{journal}{Phys. Rev.} \textbf{\bibinfo{volume}{D74}},
  \bibinfo{pages}{034508} (\bibinfo{year}{2006}), \eprint{hep-lat/0605017}.

\bibitem[{\citenamefont{Alexandrou
  et~al.}(2019{\natexlab{b}})\citenamefont{Alexandrou, Bacchio, Constantinou,
  Hadjiyiannakou, Jansen, Koutsou, and Vaquero
  Aviles-Casco}}]{Alexandrou:2019brg}
\bibinfo{author}{\bibfnamefont{C.}~\bibnamefont{Alexandrou}},
  \bibinfo{author}{\bibfnamefont{S.}~\bibnamefont{Bacchio}},
  \bibinfo{author}{\bibfnamefont{M.}~\bibnamefont{Constantinou}},
  \bibinfo{author}{\bibfnamefont{K.}~\bibnamefont{Hadjiyiannakou}},
  \bibinfo{author}{\bibfnamefont{K.}~\bibnamefont{Jansen}},
  \bibinfo{author}{\bibfnamefont{G.}~\bibnamefont{Koutsou}}, \bibnamefont{and}
  \bibinfo{author}{\bibfnamefont{A.}~\bibnamefont{Vaquero Aviles-Casco}}
  (\bibinfo{year}{2019}{\natexlab{b}}), \eprint{1909.00485}.

\bibitem[{\citenamefont{Martinelli and Sachrajda}(1987)}]{Martinelli:1987zd}
\bibinfo{author}{\bibfnamefont{G.}~\bibnamefont{Martinelli}} \bibnamefont{and}
  \bibinfo{author}{\bibfnamefont{C.~T.} \bibnamefont{Sachrajda}},
  \bibinfo{journal}{Phys. Lett.} \textbf{\bibinfo{volume}{B196}},
  \bibinfo{pages}{184} (\bibinfo{year}{1987}).

\bibitem[{\citenamefont{Martinelli et~al.}(1995)\citenamefont{Martinelli,
  Pittori, Sachrajda, Testa, and Vladikas}}]{Martinelli:1994ty}
\bibinfo{author}{\bibfnamefont{G.}~\bibnamefont{Martinelli}},
  \bibinfo{author}{\bibfnamefont{C.}~\bibnamefont{Pittori}},
  \bibinfo{author}{\bibfnamefont{C.~T.} \bibnamefont{Sachrajda}},
  \bibinfo{author}{\bibfnamefont{M.}~\bibnamefont{Testa}}, \bibnamefont{and}
  \bibinfo{author}{\bibfnamefont{A.}~\bibnamefont{Vladikas}},
  \bibinfo{journal}{Nucl. Phys.} \textbf{\bibinfo{volume}{B445}},
  \bibinfo{pages}{81} (\bibinfo{year}{1995}), \eprint{hep-lat/9411010}.

\bibitem[{\citenamefont{G{\"o}ckeler et~al.}(1999)\citenamefont{G{\"o}ckeler,
  Horsley, Oelrich, Perlt, Petters, Rakow, Sch{\"a}fer, Schierholz, and
  Schiller}}]{Gockeler:1998ye}
\bibinfo{author}{\bibfnamefont{M.}~\bibnamefont{G{\"o}ckeler}},
  \bibinfo{author}{\bibfnamefont{R.}~\bibnamefont{Horsley}},
  \bibinfo{author}{\bibfnamefont{H.}~\bibnamefont{Oelrich}},
  \bibinfo{author}{\bibfnamefont{H.}~\bibnamefont{Perlt}},
  \bibinfo{author}{\bibfnamefont{D.}~\bibnamefont{Petters}},
  \bibinfo{author}{\bibfnamefont{P.~E.~L.} \bibnamefont{Rakow}},
  \bibinfo{author}{\bibfnamefont{A.}~\bibnamefont{Sch{\"a}fer}},
  \bibinfo{author}{\bibfnamefont{G.}~\bibnamefont{Schierholz}},
  \bibnamefont{and} \bibinfo{author}{\bibfnamefont{A.}~\bibnamefont{Schiller}},
  \bibinfo{journal}{Nucl. Phys.} \textbf{\bibinfo{volume}{B544}},
  \bibinfo{pages}{699} (\bibinfo{year}{1999}), \eprint{hep-lat/9807044}.

\bibitem[{\citenamefont{Alexandrou
  et~al.}(2011{\natexlab{b}})\citenamefont{Alexandrou, Constantinou, Korzec,
  Panagopoulos, and Stylianou}}]{Alexandrou:2010me}
\bibinfo{author}{\bibfnamefont{C.}~\bibnamefont{Alexandrou}},
  \bibinfo{author}{\bibfnamefont{M.}~\bibnamefont{Constantinou}},
  \bibinfo{author}{\bibfnamefont{T.}~\bibnamefont{Korzec}},
  \bibinfo{author}{\bibfnamefont{H.}~\bibnamefont{Panagopoulos}},
  \bibnamefont{and} \bibinfo{author}{\bibfnamefont{F.}~\bibnamefont{Stylianou}}
  (\bibinfo{collaboration}{ETM Collaboration}), \bibinfo{journal}{Phys. Rev.}
  \textbf{\bibinfo{volume}{D83}}, \bibinfo{pages}{014503}
  (\bibinfo{year}{2011}{\natexlab{b}}), \eprint{arXiv:1006.1920}.

\bibitem[{\citenamefont{Alexandrou
  et~al.}(2012{\natexlab{b}})\citenamefont{Alexandrou, Constantinou, Korzec,
  Panagopoulos, and Stylianou}}]{Alexandrou:2012mt}
\bibinfo{author}{\bibfnamefont{C.}~\bibnamefont{Alexandrou}},
  \bibinfo{author}{\bibfnamefont{M.}~\bibnamefont{Constantinou}},
  \bibinfo{author}{\bibfnamefont{T.}~\bibnamefont{Korzec}},
  \bibinfo{author}{\bibfnamefont{H.}~\bibnamefont{Panagopoulos}},
  \bibnamefont{and}
  \bibinfo{author}{\bibfnamefont{F.}~\bibnamefont{Stylianou}},
  \bibinfo{journal}{Phys.Rev.} \textbf{\bibinfo{volume}{D86}},
  \bibinfo{pages}{014505} (\bibinfo{year}{2012}{\natexlab{b}}),
  \eprint{[arXiv:1201.5025]}.

\bibitem[{\citenamefont{Alexandrou et~al.}(2017)\citenamefont{Alexandrou,
  Constantinou, and Panagopoulos}}]{Alexandrou:2015sea}
\bibinfo{author}{\bibfnamefont{C.}~\bibnamefont{Alexandrou}},
  \bibinfo{author}{\bibfnamefont{M.}~\bibnamefont{Constantinou}},
  \bibnamefont{and}
  \bibinfo{author}{\bibfnamefont{H.}~\bibnamefont{Panagopoulos}}
  (\bibinfo{collaboration}{ETM}), \bibinfo{journal}{Phys. Rev.}
  \textbf{\bibinfo{volume}{D95}}, \bibinfo{pages}{034505}
  (\bibinfo{year}{2017}), \eprint{1509.00213}.

\bibitem[{\citenamefont{Constantinou et~al.}(2010)}]{Constantinou:2010gr}
\bibinfo{author}{\bibfnamefont{M.}~\bibnamefont{Constantinou}}
  \bibnamefont{et~al.} (\bibinfo{collaboration}{ETM}), \bibinfo{journal}{JHEP}
  \textbf{\bibinfo{volume}{08}}, \bibinfo{pages}{068} (\bibinfo{year}{2010}),
  \eprint{1004.1115}.

\bibitem[{\citenamefont{Constantinou et~al.}(2015)\citenamefont{Constantinou,
  Horsley, Panagopoulos, Perlt, Rakow, Schierholz, Schiller, and
  Zanotti}}]{Constantinou:2014fka}
\bibinfo{author}{\bibfnamefont{M.}~\bibnamefont{Constantinou}},
  \bibinfo{author}{\bibfnamefont{R.}~\bibnamefont{Horsley}},
  \bibinfo{author}{\bibfnamefont{H.}~\bibnamefont{Panagopoulos}},
  \bibinfo{author}{\bibfnamefont{H.}~\bibnamefont{Perlt}},
  \bibinfo{author}{\bibfnamefont{P.~E.~L.} \bibnamefont{Rakow}},
  \bibinfo{author}{\bibfnamefont{G.}~\bibnamefont{Schierholz}},
  \bibinfo{author}{\bibfnamefont{A.}~\bibnamefont{Schiller}}, \bibnamefont{and}
  \bibinfo{author}{\bibfnamefont{J.~M.} \bibnamefont{Zanotti}},
  \bibinfo{journal}{Phys. Rev.} \textbf{\bibinfo{volume}{D91}},
  \bibinfo{pages}{014502} (\bibinfo{year}{2015}), \eprint{1408.6047}.

\bibitem[{\citenamefont{Bhattacharya et~al.}(2011)\citenamefont{Bhattacharya,
  Hill, and Paz}}]{Bhattacharya:2011ah}
\bibinfo{author}{\bibfnamefont{B.}~\bibnamefont{Bhattacharya}},
  \bibinfo{author}{\bibfnamefont{R.~J.} \bibnamefont{Hill}}, \bibnamefont{and}
  \bibinfo{author}{\bibfnamefont{G.}~\bibnamefont{Paz}},
  \bibinfo{journal}{Phys. Rev.} \textbf{\bibinfo{volume}{D84}},
  \bibinfo{pages}{073006} (\bibinfo{year}{2011}), \eprint{1108.0423}.

\bibitem[{\citenamefont{Hill and Paz}(2011)}]{Hill:2011wy}
\bibinfo{author}{\bibfnamefont{R.~J.} \bibnamefont{Hill}} \bibnamefont{and}
  \bibinfo{author}{\bibfnamefont{G.}~\bibnamefont{Paz}},
  \bibinfo{journal}{Phys. Rev. Lett.} \textbf{\bibinfo{volume}{107}},
  \bibinfo{pages}{160402} (\bibinfo{year}{2011}), \eprint{1103.4617}.

\bibitem[{\citenamefont{Hill and Paz}(2010)}]{Hill:2010yb}
\bibinfo{author}{\bibfnamefont{R.~J.} \bibnamefont{Hill}} \bibnamefont{and}
  \bibinfo{author}{\bibfnamefont{G.}~\bibnamefont{Paz}},
  \bibinfo{journal}{Phys. Rev.} \textbf{\bibinfo{volume}{D82}},
  \bibinfo{pages}{113005} (\bibinfo{year}{2010}), \eprint{1008.4619}.

\bibitem[{\citenamefont{Djukanovic et~al.}(2019)\citenamefont{Djukanovic,
  Ottnad, Wilhelm, and Wittig}}]{Djukanovic:2019jtp}
\bibinfo{author}{\bibfnamefont{D.}~\bibnamefont{Djukanovic}},
  \bibinfo{author}{\bibfnamefont{K.}~\bibnamefont{Ottnad}},
  \bibinfo{author}{\bibfnamefont{J.}~\bibnamefont{Wilhelm}}, \bibnamefont{and}
  \bibinfo{author}{\bibfnamefont{H.}~\bibnamefont{Wittig}}
  (\bibinfo{year}{2019}), \eprint{1903.12566}.

\bibitem[{\citenamefont{Young et~al.}(2006)\citenamefont{Young, Roche, Carlini,
  and Thomas}}]{Young:2006jc}
\bibinfo{author}{\bibfnamefont{R.~D.} \bibnamefont{Young}},
  \bibinfo{author}{\bibfnamefont{J.}~\bibnamefont{Roche}},
  \bibinfo{author}{\bibfnamefont{R.~D.} \bibnamefont{Carlini}},
  \bibnamefont{and} \bibinfo{author}{\bibfnamefont{A.~W.}
  \bibnamefont{Thomas}}, \bibinfo{journal}{Phys. Rev. Lett.}
  \textbf{\bibinfo{volume}{97}}, \bibinfo{pages}{102002}
  (\bibinfo{year}{2006}), \eprint{nucl-ex/0604010}.

\bibitem[{\citenamefont{Liu et~al.}(2007)\citenamefont{Liu, McKeown, and
  Ramsey-Musolf}}]{Liu:2007yi}
\bibinfo{author}{\bibfnamefont{J.}~\bibnamefont{Liu}},
  \bibinfo{author}{\bibfnamefont{R.~D.} \bibnamefont{McKeown}},
  \bibnamefont{and} \bibinfo{author}{\bibfnamefont{M.~J.}
  \bibnamefont{Ramsey-Musolf}}, \bibinfo{journal}{Phys. Rev.}
  \textbf{\bibinfo{volume}{C76}}, \bibinfo{pages}{025202}
  (\bibinfo{year}{2007}), \eprint{0706.0226}.

\bibitem[{\citenamefont{Gonzalez-Jimenez
  et~al.}(2013)\citenamefont{Gonzalez-Jimenez, Caballero, and
  Donnelly}}]{GonzalezJimenez:2011fq}
\bibinfo{author}{\bibfnamefont{R.}~\bibnamefont{Gonzalez-Jimenez}},
  \bibinfo{author}{\bibfnamefont{J.~A.} \bibnamefont{Caballero}},
  \bibnamefont{and} \bibinfo{author}{\bibfnamefont{T.~W.}
  \bibnamefont{Donnelly}}, \bibinfo{journal}{Phys. Rept.}
  \textbf{\bibinfo{volume}{524}}, \bibinfo{pages}{1} (\bibinfo{year}{2013}),
  \eprint{1111.6918}.

\bibitem[{\citenamefont{Zhao}(2017)}]{Zhao:2017xej}
\bibinfo{author}{\bibfnamefont{Y.~X.} \bibnamefont{Zhao}}
  (\bibinfo{collaboration}{SoLID}), in \emph{\bibinfo{booktitle}{{22nd
  International Symposium on Spin Physics (SPIN 2016) Urbana, IL, USA,
  September 25-30, 2016}}} (\bibinfo{year}{2017}), \eprint{1701.02780}.

\bibitem[{\citenamefont{Benesch et~al.}(2014)}]{Benesch:2014bas}
\bibinfo{author}{\bibfnamefont{J.}~\bibnamefont{Benesch}} \bibnamefont{et~al.}
  (\bibinfo{collaboration}{MOLLER}) (\bibinfo{year}{2014}), \eprint{1411.4088}.

\bibitem[{\citenamefont{Armstrong et~al.}(2012)}]{Armstrong:2012ps}
\bibinfo{author}{\bibfnamefont{D.~S.} \bibnamefont{Armstrong}}
  \bibnamefont{et~al.} (\bibinfo{year}{2012}), \eprint{1202.1255}.

\bibitem[{\citenamefont{Becker et~al.}(2018)}]{Becker:2018ggl}
\bibinfo{author}{\bibfnamefont{D.}~\bibnamefont{Becker}} \bibnamefont{et~al.}
  (\bibinfo{year}{2018}), \eprint{1802.04759}.

\end{thebibliography}
\end{document}